# Relativistic and gravitational transformations in electrochemistry and nuclear magnetic resonance spectroscopy

Mirza Wasif Baig*

*** Corresponding author:** E-mail: wasifahmed797@gmail.com

## Abstract

A relativistic transformation of the electrode potential has been derived to account for time dilation effects in electrode processes. This newly formulated Lorentz transformation is interpreted in terms of the generation of spin-2 boson gravitons originating from the fusion of spin-1 virtual photons, which subsequently escape into higher dimensions. Gravitational transformations of the electrode potential have also been derived, explaining the observed decrease in cell potential under stronger gravitational fields. The reduction in electrode potential near a gravitational source is attributed to a greater flux of gravitons escaping into higher dimensions in stronger gravitational fields compared to weaker ones. Similarly, the potential energy associated with the spin of magnetically active nuclei in an applied magnetic field, as observed in nuclear magnetic resonance (NMR) spectroscopy, is shown to be Lorentz-variant. This provides a mathematical demonstration that the Hamiltonian describing the energy of such nuclei is also Lorentz-variant. The relativistic and gravitational transformations of both the electrode potential and the spin-related potential energy in magnetic fields are shown to be analogous.

## 1. Introduction

Electrochemistry is an interdisciplinary science which comes from the marriage of electrical physics and chemistry [1]. Majorly in electrochemistry, redox reactions (oxidation and reduction) occurring at the interface of electrode and electrolyte are studied which occurs in the liquid phase. Electrochemical reactions occurring at the surface of the electrode can be broadly classified in two categories non-spontaneous and spontaneous redox reactions. Non-spontaneous redox reactions are carried out by passing an electric current through the cell by applying at least threshold potential across the electrodes of the cell while spontaneous redox reactions govern on their own when electrodes are brought in contact with a suitable electrolyte which generates a potential difference across the electrodes that lead to a flow of electric current through the external circuit [2]. NMR spectroscopy involves transition of the orientation of nuclear spins in presence of applied magnetic field within NMR spectrometer. Quantum mechanical states of nuclear spins interact with the magnetic fields and transitions between these quantum states are induced when nuclei are being irradiated with electromagnetic radiation [3]. Energy spacing between permitted quantum levels of molecules has been mathematically proved to be Lorentz variant and reported to decrease for moving observer [4]. Thus, it is a natural question to ask how energy spacing between quantum levels of nuclear spins will behave for moving observer. To answer this question in present work NMR spectroscopy from an eye of moving observer has been explained. Relativistic thermodynamics and kinetics have been formulated to explain time dilation for chemical reactions in a moving frame of reference [4-5]. Relativistic time dilation is one of the basic foundations of the special theory of relativity [6]. In the relativistic theory of chemical kinetics Lorentz transformation of rate constant and thermodynamic state functions have been formulated which are compatible with time dilation phenomenon, which is an experimentally verified fact [7]. Similarly, gravitational time dilation which comes from the general theory of relativity is also an experimentally verified fact reveals that clocks tick slower near the source of gravity than at an appreciable height let say "$h$" from the surface of heavy planet [8-9]. Gravitational time dilation will thus also affect the rate of molecular processes at different gravitational fields. The effect of the strength of the gravitational field on the rates of reactions has been formulated in terms of gravitational transformations of rate constants, activation energy and thermodynamic state functions [10]. Gravitational transformation of energy spacing between permitted quantum levels of molecules has also been mathematically proved i.e., near to surface of heavy planets energy spacing between quantum levels squeezes revealing the quantumphobic nature of gravity [10]. Again, it is a natural question to ask how energy spacing between quantum levels of nuclear spins will behave in stronger and weaker gravitational fields. So, the present work is a first most attempt to explain how gravity will affect the NMR experiments. In addition, the present paper also formulates Lorentz and gravitational transformations that can explain relativistic and gravitational time dilation in the kinetics of electrochemical phenomenon. In the present paper, all the variables with subscript "$o$" refer to the inertial frame of reference $K_o$ with the observer at rest while variables with subscript u refer to the inertial frame $K_u$ with the observer moving at relative speed "$u$". All Lorentz transformations are defined in terms of a factor "$\gamma = (1 - u^2/c^2)^{-1/2}$". To explain gravitational time dilation all the variables for observer $K_o$ with radial coordinate $r$ is located at the surface of massive planet "$M$" are described by subscript "$s$" with reference to observer $K_h$ located at height "$h$" from the surface of massive planet "$M$". Their magnitude at different positions in the gravitational field of massive planet "$M$" is mathematically related with a factor defined as "$\xi = (1 - 2GM/rc^2)^{-1/2}$". When $\xi^{-1}$will appear in gravitational transformation it will be termed as β-gravitational transformation as used in previous work [10].

## 2. Theory I

To formulate a theory that can explain time dilation for the electrochemical process from an eye of moving observer, let consider following reaction occurring on the surface of an arbitrary electrode [11].

$$O + e^- \underset{k_b}{\overset{k_f}{\rightleftharpoons}} R \tag{1}$$

Let in above electrode process specie O and R is being involved in one electron transfer process. Free energy associated with above one electron electrode process for moving observer can be formulated as [11, 12]

$$\Delta G_u = -nFE_u \tag{2}$$

As free energy is proved to be Lorentz variant i.e. $\Delta G_u = \gamma^{-1}\Delta G_o$ [4, 5] it is observed to be less for moving observer. Since according to the special theory of relativity laws of physics holds good in all inertial frames. So, to hold this basic postulate of the special relativity electrode potential becomes Lorentz variant i.e.

$$E_u = \gamma^{-1}E_o \tag{3}$$

Rate constants of forward and backward reaction can be formulated in terms of Arrhenius factor and free energy of activation as [12, 13],

$$\left(k_f\right)_u = \left(A_f\right)_u \exp\left(-\left(\Delta G_{0a}^{\dagger}\right)_u / R_u T\right) \exp[-\alpha(E_u - E_u^0/R_u T) \tag{4}$$

$$(k_b)_u = (A_b)_u \exp\left(-\left(\Delta G_{0b}^{\dagger}\right)_u / R_u T\right) \exp[(1-\alpha)(E_u - E_u^0/R_u T) \tag{5}$$

For a special case $E_u = E_u^0$ rate constant for forward and backward reaction becomes equal i.e.

$$(k_0)_u = (A_b)_u \exp\left(-\left(\Delta G_{0b}^{\dagger}\right)_u / R_u T\right) = \left(A_f\right)_u \exp\left(-\left(\Delta G_{0a}^{\dagger}\right)_u / R_u T\right) \tag{6}$$

On substitution of Lorentz transformation of Arrhenius factor, ideal gas constant free activation energy in eq. (6) gives same Lorentz transformation of threshold rate constant i.e. $(k_0)_u = \gamma^{-1}(k_0)_o$. Eqs. (4) and (5) can be written in more contact form as [1, 2],

$$\left(k_f\right)_u = (k_0)_u \exp[-\alpha(E_u - E_u^0/R_u T) \tag{7}$$

$$(k_b)_u = (k_0)_u \exp[(1-\alpha)(E_u - E_u^0/R_u T) \tag{8}$$

On substitution of Lorentz transformation of electrode potential, ideal gas constant and threshold rate constant in eqs. (7) and (8) gives the same Lorentz transformation of the rate constant as formulated earlier [4]

$$\left(k_f\right)_u = \gamma^{-1}\left(k_f\right)_o \tag{9}$$

$$(k_b)_u = \gamma^{-1}(k_b)_o \quad (10)$$

Butler-Volmer relation that explains total current flow in an electrode process for moving observer can be stated as [2, 12],

$$i_u = NFA(k_0)_u[\exp[-\alpha(E_u - E_u^0/R_uT) - \exp[(1-\alpha)(E_u - E_u^0/R_uT)] \quad (11)$$

On substitution of Lorentz transformation of electrode potential, ideal gas constant and threshold rate constant in eq. (10) gives the Lorentz transformation of current as [4],

$$i_u = \gamma^{-1}i_o \quad (12)$$

So due to the drop in the electric potential for moving observer net current flow will slow down. The moving observer will find net charge transfer per unit time is less than compared to in its own frame of reference because for moving observer clock of the system at rest will tick slower.

Electrode potential associated with above mentioned one electron electrode process at stronger gravitational potential i.e. near the source of gravity can also be formulated as [2, 12],

$$E_s = -\Delta G_s/nF \quad (13)$$

Since free energy is lower at stronger gravitational field i.e., $\Delta G_s = \xi^{-1}\Delta G_h$ [10]. So, this formulates that electrode potential to have β-gravitational transformation [10] i.e.

$$E_s = \xi^{-1}E_h \quad (14)$$

Rate constants of forward and backward reaction can be formulated in terms of Arrhenius factor and free energy of activation as for system near the source of gravity i.e. at the surface of massive planet can be expressed as [12, 13],

$$\left(k_f\right)_s = \left(A_f\right)_s \exp\left(-\left(\Delta G_{0a}^{\dagger}\right)_s/R_sT\right) \exp[-\alpha(E_s - E_s^0/R_sT) \quad (15)$$

$$(k_b)_s = (A_b)_s \exp\left(-\left(\Delta G_{0b}^{\dagger}\right)_s/R_sT\right) \exp[(1-\alpha)(E_s - E_s^0/R_sT) \quad (16)$$

For a special case $E_s = E_s^0$ rate constant for forward and backward reaction becomes equal i.e.

$$(k_0)_s = (A_b)_s \exp\left(-\left(\Delta G_{0b}^{\dagger}\right)_s/R_sT\right) = \left(A_f\right)_s \exp\left(-\left(\Delta G_{0a}^{\dagger}\right)_s/R_sT\right) \quad (17)$$

On substitution of the gravitational transformation of Arrhenius factor, ideal gas constant and free activation energy in eq. (17) gives the same gravitational transformation of threshold rate constant i.e. $(k_0)_s = \xi^{-1}(k_0)_h$.

Eqs. (15) and (16) can be written in more compact form as [1, 2],

$$\left(k_f\right)_s = (k_0)_s \exp[-\alpha(E_s - E_s^0/R_sT) \quad (18)$$

$$(k_b)_s = (k_0)_s \exp[(1-\alpha)(E_s - E_s^0/R_sT) \quad (19)$$

On substitution of gravitational transformation of electrode potential, ideal gas constant and threshold rate constant in eqs. (18) and (19) gives the same β-gravitational transformation of rate constant as formulated earlier [10],

$$(k_f)_s = \xi^{-1}(k_f)_h \quad (20)$$

$$(k_b)_s = \xi^{-1}(k_b)_h \quad (21)$$

Butler-Volmer relation that explains total current flow in an electrode process at the stronger gravitational field can be stated as [12],

$$i_s = NFA(k_0)_s[\exp[-\alpha(E_s - E_s^0/R_sT) - \exp[(1-\alpha)(E_s - E_s^0/R_sT)] \quad (22)$$

On substitution of the gravitational transformation of electrode potential, ideal gas constant and threshold rate constant in eq. (22) gives the gravitational transformation of current as,

$$i_l = \xi^{-1} i_h \quad (23)$$

So, due to decrease in electric potential at the stronger gravitational field, an observer at the stronger gravitational field will find net charge transfer per unit time smaller than compared to an observer at the weaker gravitational potential (away from the surface of heavy planet) this is because of the slower ticking of clocks at the stronger gravitational field than at higher gravitational field.

## 3. Theory II

To formulate a relativistic theory for NMR spectroscopy considered a molecule AH possessing a proton H is placed in NMR spectrometer in which strength of applied magnetic field is $B_{appl}$. Potential energy experienced by this molecule in NMR spectrometer for moving observer can be mathematically expressed as [14],

$$V_u = -\Omega_u(1 - \sigma B_{appl})\hat{I}_z \quad (24)$$

Where $\Omega_u$ in eq. (23) is defined as [14]

$$\Omega_u = g_N \frac{q}{2(m_H)_u} \quad (25)$$

Using relativistic transformation of mass for the mass of proton $m_H$ [6] in eq. (25) gives following Lorentz transformation for $\Omega_u$

$$\Omega_u = \gamma^{-1}\Omega_0 \quad (26)$$

From eq. (26) Lorentz transformation for potential energy experienced by this molecule can be mathematically formulated as,

$$V_u = \gamma^{-1} V_0 \quad (27)$$

This transformation is totally consistent with that obtained for electrode potential experienced by ion given earlier by eq. (2). Hamiltonian acting on quantum mechanical wave function of proton will be described by following Schrödinger equation [14].

$$\widehat{H}_u \psi = E_u \psi \tag{28}$$

Whereas quantum mechanical Hamiltonian describing energy of protons in applied magnetic field $B_{appl}$ will be mathematically defined as [3, 14],

$$\widehat{H}_u = -\Omega_u (1 - \sigma B_{appl}) \hat{I}_z \tag{29}$$

Using Lorentz transformation of $\Omega_u$ given by eq. (26) gives following Lorentz transformation of the quantum mechanical Hamiltonian describing energy of protons.

$$\widehat{H}_u = \gamma^{-1} \widehat{H}_0 \tag{30}$$

Energy of protons described in eq. (28) will be mathematically expressed as [14]

$$E_u = -\hbar \Omega_u m_I (1 - \sigma B_{appl}) \tag{31}$$

Using Lorentz transformation of $\Omega_u$ given by eq. (26) gives the following Lorentz transformation of the energy of protons.

$$E_u = \gamma^{-1} E_0 \tag{32}$$

Using eq. (31) energy gap between two-quantum spin states of protons can be easily obtained is [14]

$$\triangle E_u = \hbar \Omega_u (1 - \sigma B_{appl}) \tag{33}$$

Using Lorentz transformation of $\Omega_u$ given by eq. (26) gives the following Lorentz transformation of the energy gap between two quantum spin-states of protons.

$$\triangle E_u = \gamma^{-1} \triangle E_0 \tag{34}$$

Eq. (34) is totally consistent with Lorentz transformation of energy spacing proved in earlier work where for moving observer spacing between quantum levels associated with rotation vibrations and electronic jumps in molecules squeeze for moving observer. Since Planks constant is Lorentz invariant thus dividing eq. (33) with Planks constant also mathematically proves that frequency [3, 14] associated with quantum levels for nuclear spin is also Lorentz variant i.e.,

$$\omega_u = \gamma^{-1} \omega_0 \tag{35}$$

Now considering a molecule $AH_AH_B$ possessing two chemically non-equivalent protons placed in NMR spectrometer in which strength of applied magnetic field is $B_{appl}$. Hamiltonian for such a system will be formulated as [14]

$$\widehat{H}_u = -\Omega_u \big(1 - \sigma_A B_{appl}\big) \hat{I}_{zA} - \Omega_u \big(1 - \sigma_B B_{appl}\big) \hat{I}_{zB} + \frac{h (J_{AB})_u}{\hbar^2} \hat{I}_{zA} \bullet \hat{I}_{zB} \tag{36}$$

As it is shown in earlier work that electronic coupling is Lorentz variant and decreases for moving observer i.e., $(H_{AB})_u = \gamma^{-1}(H_{AB})_0$ [4]. Similarly, coupling between the magnetically equivalent protons will also be Lorentz variant and thus decreases for moving observer i.e.,

$$(J_{AB})_u = \gamma^{-1}(J_{AB})_0 \tag{37}$$

Substituting Lorentz transformation of coupling between the nuclei from eq. (37) and Lorentz transformation of $\Omega_u$ from eq. (26) also gives same Lorentz transformation of Hamiltonian for this system $AH_AH_B$ as obtained earlier in Eq. (30) i.e.

$$\hat{H}_u = \gamma^{-1}\hat{H}_0 \tag{38}$$

To derive gravitational transformation for different NMR parameters again considered a molecule AH possessing a proton H is placed in NMR spectrometer operating at stronger gravitational field in which strength of applied magnetic field is $B_{appl}$. Potential energy experienced by this molecule in NMR spectrometer at the surface of heavy planet can be mathematically expressed as [14],

$$V_s = -\Omega_s(1 - \sigma B_{appl})\hat{I}_z \tag{39}$$

Where $\Omega_s$ in eq. (39) is defined as [14]

$$\Omega_s = g_N \frac{q}{2(m_H)_s} \tag{40}$$

Using $\boldsymbol{\alpha}$-gravitational transformation of mass [10] for the mass of proton $m_H$ in eq. (40) gives following gravitational transformation for $\Omega_s$

$$\Omega_s = \xi^{-1}\Omega_h \tag{41}$$

From eq. (41) gravitational transformation for potential energy experienced by this molecule can be mathematically formulated as,

$$V_s = \xi^{-1}\, V_h \tag{42}$$

This transformation is totally consistent with that obtained for electrode potential experienced by ion given earlier by eq. (13). Hamiltonian acting on quantum mechanical wave function of proton will be described by following Schrödinger equation [3, 14].

$$\hat{H}_s\psi = E_s\psi \tag{43}$$

Whereas quantum mechanical Hamiltonian describing energy of protons in applied magnetic field $B_{appl}$ will be mathematically defined as [3, 14],

$$\hat{H}_s = -\Omega_s(1 - \sigma B_{appl})\hat{I}_z \tag{44}$$

Using gravitational transformation of $\Omega_s$ given by eq. (41) gives following β-gravitational transformation of the quantum mechanical Hamiltonian describing energy of protons.

$$\hat{H}_s = \xi^{-1}\hat{H}_h \tag{45}$$

Energy of protons described in eq. (43) will be mathematically expressed as

$$E_s = -\hbar\Omega_s m_s(1 - \sigma B_{appl}) \tag{46}$$

Using gravitational transformation of $\Omega_s$ given by eq. (41) gives the following β-gravitational transformation of the energy of protons.

$$E_s = \xi^{-1} E_h \tag{47}$$

Using eq. (46) energy gap between two-quantum spin states of protons can be easily obtained as

$$\triangle E_s = \hbar\Omega_s(1 - \sigma B_{appl}) \tag{48}$$

Using gravitational transformation of $\Omega_l$ given by eq. (41) gives the following gravitational transformation of the energy gap between two quantum spin-states of protons.

$$\triangle E_s = \xi^{-1} \triangle E_h \tag{49}$$

Eq. (49) is totally consistent with gravitational transformation of energy spacing [10] proved in the earlier work where at stronger gravitational potential spacing between quantum levels associated with rotation vibrations and electronic jumps in molecules is smaller than at weaker gravitational field. Since magnitude of Planks constant is unaffected by gravitational field thus dividing eq. (33) with Planks constant also mathematically proves that frequency [3, 14] associated with quantum levels for nuclear spin has following β-gravitational transformation i.e.,

$$\omega_s = \xi^{-1}\, \omega_h \tag{50}$$

Now considering a molecule $AH_AH_B$ possessing two chemically non-equivalent protons placed in NMR spectrometer in which strength of applied magnetic field is $B_{appl}$. Hamiltonian for such a system at stronger gravitational field will be formulated as [14]

$$\widehat{H}_s = -\Omega_s\big(1 - \sigma_A B_{appl}\big)\hat{I}_{zA} - \Omega_l\big(1 - \sigma_B B_{appl}\big)\hat{I}_{zB} + \frac{h(J_{AB})_l}{\hbar^2}\hat{I}_{zA} \bullet \hat{I}_{zB} \tag{51}$$

As it is shown in earlier work that electronic coupling undergoes gravitational transformations [10] and decreases at stronger gravitational field i.e., $(H_{AB})_s = \xi^{-1}(H_{AB})_h$. Similarly, coupling between the magnetically equivalent protons will also have β-gravitational transformation and thus decreases at stronger gravitational field i.e.,

$$(J_{AB})_s = \xi^{-1}(J_{AB})_h \tag{52}$$

Substituting gravitational transformation of coupling between the nuclei from eq. (52) and gravitational transformation of $\Omega_s$ from eq. (41) also gives same β-gravitational transformation of the Hamiltonian for this system $AH_AH_B$ as obtained earlier in Eq. (45) i.e.

$$\widehat{H}_s = \xi^{-1}\widehat{H}_h \tag{53}$$

4. **Discussion**

Migration of ions toward their respective electrodes occurs due to the electrostatic force of attraction between the electrode and the electrolyte. The potential applied externally to the electrode, or the potential developed at the electrode upon contact with the electrolyte, renders it either positively or negatively charged. This charge attracts oppositely charged ions toward the electrode, where they undergo the corresponding redox reactions. According to Eq. (2), if the cell potential is observed to decrease for a moving observer, this implies that the magnitude of the electrostatic force of attraction between the ions and the electrode also decreases for that observer. To explain this phenomenon, we must draw upon certain fundamental concepts from particle physics [15]. For every fundamental force in nature, there exists a corresponding particle that mediates the interaction [16]. In the case of the electrostatic force, the mediator is the photon, a spin-1 boson [17, 18]. The decrease in potential observed by a moving observer can be interpreted in terms of the interconversion of virtual photons into gravitons. The gravitational force is mediated by a hypothetical spin-2 boson known as the graviton, which may escape into higher dimensions [19, 20]. Gravitation is considered a weak force due to a distinctive property attributed to gravitons: they are associated with closed strings, allowing them to propagate into higher-dimensional space and leak from the brane [21, 22]. The possibility of gravitons escaping into any of the $n$ extra dimensions in a total $4 + n$ dimensional framework is a well-known theoretical concept frequently discussed in the literature [23, 25]. D. Boccaletti *et al.* have described the production of gravitons via the fusion of photons in the presence of either a gravitational or an electromagnetic field [26, 27]. The generation of gravitons during photon–photon, electron–photon, and nucleus–nucleus collisions at the TeV scale has also been predicted in the literature [28, 30]. The potential between an electrode and ions is responsible for the electrostatic force that attracts ions toward the electrode. In a moving frame, a decrease in this potential leads to a reduction in the magnitude of the electromotive force between the ions and the electrode. For a moving observer, the drop in potential and the weakening of the electromotive force occur due to the interconversion of virtual photons (which mediate the electromotive force between ions and the electrode) into gravitons. The potential drop observed by a moving observer, at any temperature, results from the fusion of virtual photons producing gravitons, which subsequently escape into one of the $n$ higher dimensions of a total $4 + n$ dimensional space. Specifically, two virtual photons combine to form a graviton, which escapes into higher dimensions. A diagrammatic representation of virtual photons $(\gamma^1)$ combining to produce a graviton $(g^2)$escaping into a higher dimension is shown in Fig. 1.

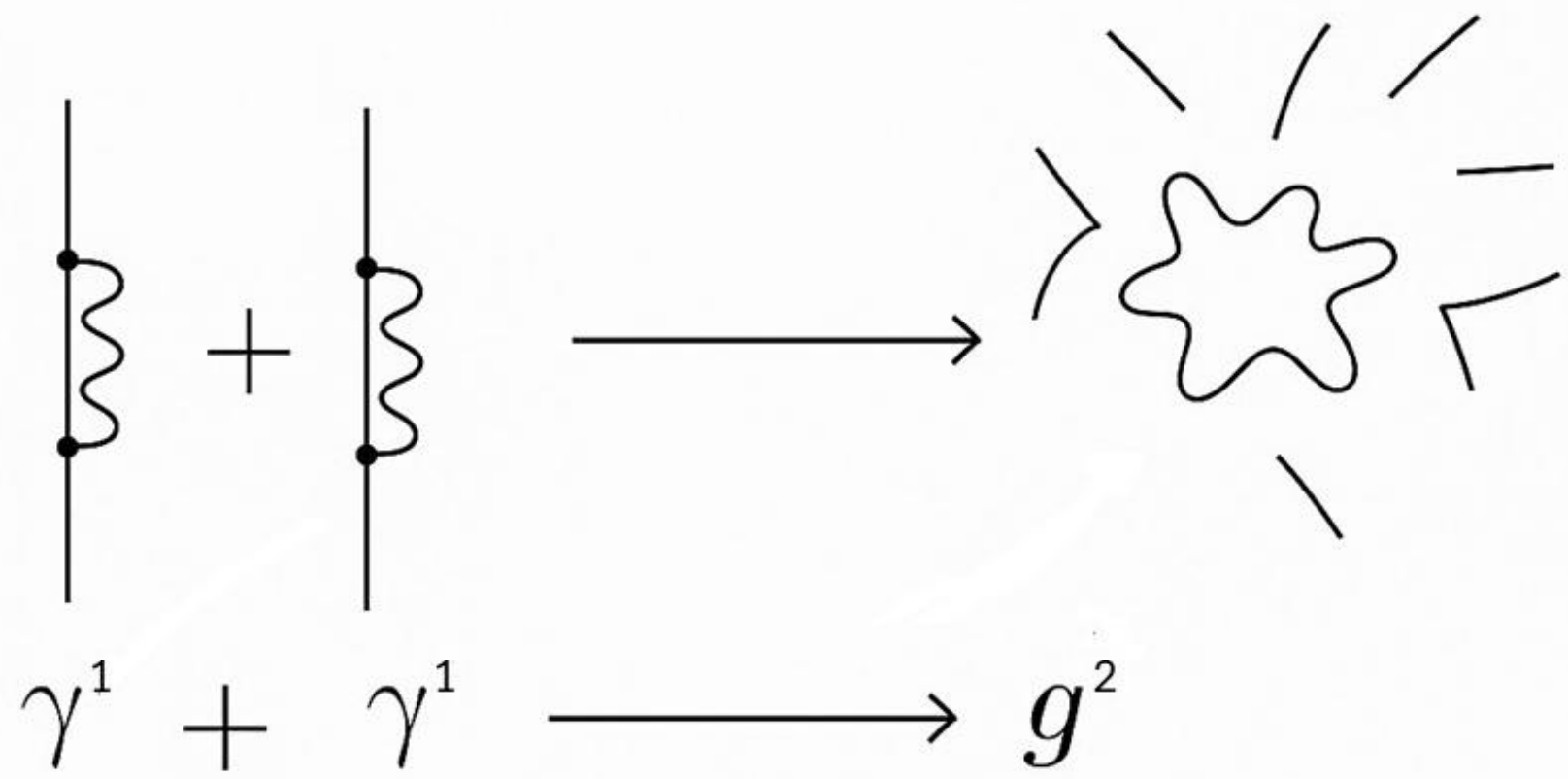

**Figure 1:** Diagrammatic representation of production of a graviton from virtual photons.

The conversion of a photon into a graviton, and the reverse process of graviton decay into a photon, have already been reported in the literature [31–33]. In the present theory, it is postulated, for the first time, that for a moving observer at any arbitrary energy scale and temperature, two spin-1 bosons ("photons") can combine to produce a spin-2 boson ("graviton"), which then disappears into one of the possible *n* higher dimensions of a total 4 + *n* dimensional space. This interconversion of photons into gravitons leads to a decrease in the potential of the cell and the electrode for the moving observer. In this frame, some of the photons mediating the electrostatic force of attraction between ions and the electrode combine to form gravitons, which escape into any of the *n* higher dimensions of the 4 + *n* dimensional space. This mechanism results in the Lorentz transformation of the cell and electrode potential for the moving observer, expressed as. $\left(E_{cell/elect}\right)_u = \gamma^{-1}\left(E_{cell/elect}\right)_0$. Similarly, the spins of magnetically active nuclei interact with the applied magnetic field in an NMR spectrometer. This interaction is electromagnetic in nature. In a moving frame, due to the effective increase in the mass of electrons and nucleons, the gyromagnetic ratio decreases, i.e., $\Omega_u = \gamma^{-1}\Omega_0$, causing the spins to interact less strongly with the applied magnetic field. The reduction in the spin–field interaction is also a consequence of the enhanced escape of gravitons into higher dimensions in the moving frame, resulting in the Lorentz transformation of the potential energy of magnetically active nuclei under the applied magnetic field, i.e., $V_u = \gamma^{-1}V_0$. The Lorentz transformations of electrode potential and nuclear potential energy are analogous to one another. Since relativity is always incorporated within the Hamiltonian [34–35] in quantum-mechanical equations, the relativistic transformation of electric potential renders the Hamiltonian Lorentz variant. According to Eq. (38), relativistic Hamiltonians [36–37] developed in relativistic quantum chemistry should be complete only if they are Lorentz variant; otherwise, they are incomplete, i.e., unable to account for relativistic time-dilation effects. Similarly, according to Eq. (45), relativistic Hamiltonians should include β-gravitational transformations; otherwise, they are incomplete and cannot explain gravitational time-dilation effects. Kohn–Sham–based density functional theory (DFT) methods [38–40], which are commonly used to compute the energetics of electronic energy levels in molecular systems [41], explicitly include the Kohn–Sham potential—a sum of the external, Hartree, and exchange–correlation potentials. According to Eq. (27), these methods are Lorentz variant and, according to Eq. (42), they also exhibit β-gravitational transformations. Even the Dirac Hamiltonian [42, 43] in the relativistic wave equation is incomplete, as it does not account for the Lamb shift [44], parity-violation effects [45], or time-dilation effects. Both the Pauli Hamiltonian [46], derived from the non-relativistic limit of the Dirac equation and accounting for spin–orbit and magnetic interactions, and the Kroll Hamiltonian [47], which includes additional relativistic and QED perturbative corrections to the non-relativistic Coulomb Hamiltonian, are used to compute spin–orbit couplings (SOCs) [48]; however, both are also incomplete. Spin–orbit couplings (SOCs), analogous to electronic couplings, are expected to follow Lorentz [4] and β-gravitational transformations [10], thereby giving rise to time-dilation effects in intersystem crossing reactions [49], as theoretically demonstrated for other chemical processes [4, 10]. Accordingly, a complete Hamiltonian must rigorously account for scalar relativistic effects, spin–orbit couplings, the Lamb shift, parity-violating interactions, and both relativistic and gravitational time-dilation effects. The β-gravitational transformation of cell potential, shown in Eq. (14), reveals that in a stronger gravitational field—i.e., near the source of gravity—the electrostatic force of attraction between ions and the electrode decreases in magnitude, resulting in a drop in cell potential. This decrease can be explained in terms of the larger flux of gravitons escaping into higher dimensions at stronger gravitational fields, analogous to the Lorentz transformation of cell potential. The α-gravitational transformation of mass can similarly be understood as arising from virtual photons combining to produce gravitons, which then escape into higher dimensions. A greater flux of gravitons escaping into higher dimensions leads to an increase in gravitational mass at stronger gravitational fields. Atoms and molecules consist of positively charged protons in the nuclei and negatively charged electrons surrounding them. Two forces exist between revolving electrons and protons: the electrostatic force of attraction, due to their charges, mediated by the exchange of virtual photons [16]; and the gravitational force, due to their masses, mediated by the exchange of gravitons [23]. In a stronger gravitational field, some virtual photons combine to produce gravitons that escape into higher dimensions, whereas at weaker gravitational fields, fewer gravitons escape. Consequently, the flux of gravitons escaping into higher dimensions is greater at stronger gravitational fields. The increase in graviton flux at stronger gravitational fields arises from the fusion of virtual photons mediating the electrostatic attraction between electrons and protons, which weakens the electrostatic force. Similarly, the virtual photons mediating the electrostatic force between the electrode and electrolyte decrease at stronger gravitational fields because some photons combine to form gravitons that escape into higher dimensions. As a result, some of the photons mediating the electrostatic attraction between ions and electrodes are converted into gravitons, weakening the electrostatic force and leading to the gravitational transformation of cell and electrode potential $E_s = \xi^{-1}E_h$. Similarly, the gravitational transformation of the potential energy of nucleons in an applied magnetic field is given by $V_s = \xi^{-1}V_h$. This occurs because the effective mass of nucleons increases in stronger gravitational fields, which decreases the gyromagnetic ratio, i.e., $\Omega_s = \xi^{-1}\Omega_h$, causing weaker interaction with the applied magnetic field. The reduction in spin–field interaction is also a consequence of the enhanced escape of gravitons into higher dimensions in stronger gravitational fields. Since relativity is always incorporated within the Hamiltonian [34–35] of quantum-mechanical equations, gravitational effects should likewise be included. Consequently, the β-gravitational transformation of electric potential renders the Hamiltonian β-gravitationally variant, as shown in Eq. (53). The gravitational transformations of electrode potential and the potential energy of nuclei are analogous. Experimental verification of the present work can be conducted by studying very simple H+ $H_2$ and D+ $H_2$ chemical reactions [50] in moving frames at relativistic speeds. Since it is difficult to accelerate neutral atoms, $H^+$ and $D^+$ ions can be accelerated to relativistic speeds and then allowed to undergo reactive collisions with $H_2$ molecules. At the instant of reactive collision, the energy released will be lower than in the rest frame due to the escape of gravitons into higher dimensions. The present work thus attempts to uncover the relationship between thermodynamics, quantum mechanics, and relativity. Lucia et al. also proposed a thermodynamic definition of time and established a link between quantum mechanics, relativity, and thermodynamics [51]. However, they again claimed a Lorentz transformation of thermodynamic temperature, which is incorrect. Mareš et al. have reported that the strictly local phenomenological temperature $T$, defined from an empirical temperature scale based on a sequence of fixed thermometric reference points, should remain invariant under Lorentz transformations [52, 53]. They also note that full equivalence between thermodynamics and statistical physics is not generally feasible, although this equivalence was successfully achieved in the relativistic theory of reaction rates [4]. Ming et al. investigated the fundamental connection between quantum mechanics and thermodynamics by establishing a no-go theorem linking the irreversible growth of correlations, entropy increase, approach to equilibrium, and decoherence as manifestations of a single underlying principle [54]. Since the second law of thermodynamics connects the arrow of time to the monotonic increase of entropy, time dilation implies that entropy must be Lorentz variant [5]. This is only possible if heat is Lorentz variant [5] while temperature remains Lorentz invariant [52, 53], a conclusion that challenges earlier formulations of Lorentz-transformed temperature [51]. Instanton rate theory [55–59], which explains tunneling rates of chemical reactions below the crossover temperature, relates the rate constant for tunneling through a barrier to the period of a classical trajectory on the inverted potential energy surface, with period $\tau = nk_BT/\hbar$ [60], where "n" is 1 or 2 depending on the definition of the crossover temperature [61]. This is essentially the same as the "periodic time of oscillation" described by Lucia et al. [51]. Since the critical temperature marking the transition from Maxwell–Boltzmann to Fermi–Dirac or Bose–Einstein statistics is invariant across inertial frames, the crossover temperature [61–63], below which tunneling is important in chemical reactions, should also remain the same for all observers, regardless of relative speed. This can be mathematically verified by substituting the Lorentz-transformed frequency and Boltzmann constant into the crossover temperature relation, $T_c = \hbar\omega_B/2\pi k_B$. Thus, relativistic [4] and gravitational [10] transformations of the Boltzmann constant map the period of a classical trajectory on an inverted potential energy surface onto Einstein's time-dilation relations. This explains time dilation in tunneling rates and indicates a deeper connection between quantum mechanics and relativity through thermodynamic principles. It is worth noting that, to date, there is no unique quantum transition state theory (QTST) [64–68] due to the uncertainty principle and other factors [69]. Semiclassical instanton rate theory outperforms QTST in regimes dominated by quantum tunneling, including low temperatures, high or wide potential barriers, and reaction pathways involving significant quantum barrier penetration [70]. The crossover temperature [61–63], defined as the temperature at which the instanton energy equals the barrier height, can also be interpreted thermodynamically, again suggesting a hidden connection between quantum mechanics and relativity. Maldacena, Shenker, and Stanford proposed a fundamental quantum bound for the rate of information scrambling, given by the Lyapunov exponent $\lambda_L = 2\pi k_B T/\hbar$ [71]. For a quantum system approaching the event horizon of a black hole, the spread of quantum information (information scrambling [72]) becomes increasingly difficult and approaches zero at the event horizon due to the gravitational transformation of the Boltzmann constant. Information scrambling is characterized in terms of the growth of Heisenberg operators, linking it to the concept of rewinding time [72]. In QTST, the transmission coefficient, which accounts for quantum tunneling and recrossing effects, is mathematically evaluated as the trace of the product of the flux operator, the Boltzmann operator, and the projection operator [73, 74]; the flux and projection operators are treated as Heisenberg operators. Tunneling in various physical systems [75], particularly in chemical reactions [76], can act as an information scrambler [77]. The gravitational transformation of the Boltzmann constant implies that, for a molecular system, the rate of information scrambling is suppressed, which in turn reduces tunneling. It is important to note that the gravitational transformation of the Boltzmann constant applies only to a system approaching a massive body, such as a black hole, and cannot be used to evaluate the entropy of a black hole [78, 79]. In the present work, black holes nature's ultimate information scramblers [79] form when the mass of a massive star collapses under gravity, compressing all matter into a spacetime singularity and converting virtual spin-1 particles, which mediate electromagnetic and strong interactions, into spin-2 gravitons. These gravitons may propagate into extra dimensions, encoding quantum information at the event horizon, while the black hole funnels all infalling matter toward the singularity. A nitrogen molecule ($N_2$) and a carbon monoxide molecule (CO), despite having identical atomic mass units, would contribute the same mass to a black hole's singularity but generate distinct graviton distributions on the event horizon, preserving information [80]. By virtue of the Hohenberg–Kohn theorem [81], the ground-state electron densities of CO and $N_2$ uniquely determine the corresponding external potentials $v_{CO}(r)$and $v_{N_2}(r)$, up to an additive constant. By analogy, the present work suggests that distinct molecular electron densities correspond to distinct numbers of graviton depositions upon crossing an event horizon, and that their residual entropy likewise drops to zero. The spacetime singularity arises from the convergence of all three spatial dimensions and time, while the event horizon and interior of the black hole are hypothesized to involve the remaining extra dimensions. Since thermal radiation [82] and heat transfer [83] in black holes are deeply connected to quantum field theory, general relativity, and thermodynamics, the present work opens new horizons for exploring black hole physics.

**Application**

**4.1. Daniel cell**

Daniel cell is one of the earliest electro voltaic cells in which Zn(0) is oxidized to Zn(II) at the anode, Cu(II) is reduced to Cu(0) at the cathode, i.e. $Zn + Cu^{2+} \rightarrow Zn^{2+} + Cu$ [11]. To demonstrate the quantitative effect of the relativistic time dilation, consider the moving observer monitors cell potential of at Daniel cell speed of $2\times10^{8} ms^{-1}$ then cell potential will be observed to decrease as shown in Table 1.

| **Table 1:** Comparison of cell potential for moving and stationary observer at 298 K | |
|---|---|
| Net cell potential for stationary observer (V) | Net cell potential for moving observer (V) |
| 1.10 | 0.819[a] |
| [a] This value has been developed from eq. (3) | |

Now to demonstrate the quantitative effect of the gravitational field strength on cell potential, consider two Daniel cells one operating at surface of earth other operating in laboratory named "$CZ$" [10] constructed at a distance "$h$" equal to 939 km away from the event horizon of black hole M87 [84]. Temperature and pressure at surface of earth and laboratory named "$CZ$" are kept constant so that solely the effect of gravity on cell potential of can be considered. Since gravitational field strength at the surface of the Earth is much weaker than the gravitational field strength at a distance "$h$" of 939 km away from the event horizon of black hole M87 [84]. Thus, gravitational transformation of cell potential of Daniel cell can be mathematically expressed as;

$$E_{M87} = E_{earth}\left(\sqrt{1-2GM87/(R_{M87}+h)\,c^{2}}\right)/\left(\sqrt{1-2GM_{earth}/R_{earth}c^{2}}\right) \tag{54}$$

The effect of gravity on the cell potential of Daniel cell operating in laboratory named "$CZ$" located at a distance "$h$" of 939 km away from the event horizon of black hole M87 [84] is elaborated in Table 2.

| **Table 2:** Comparison of cell potentials at surface of earth and laboratory "$CZ$" [10] 939 km away from the event horizon of black hole M87 [84] | |
|---|---|
| Net cell potential at surface of earth (V) | Net cell potential at laboratory "$CZ$" (V) |
| 1.10 | 0.11[b] |
| [b] This value has been developed from eq. (54) | |

## 5. Conclusion

There are two main approaches to merging quantum mechanics with the general theory of relativity: one aims to develop a theory of quantum gravity [85, 86], and the other seeks to gravitize quantum mechanics [87, 88]. The present theory suggests that a connection between quantum mechanics and the general theory of relativity may be established through thermodynamics, thereby supporting the idea of treating gravity as an entropic force [89]. Lorentz and gravitational transformations of free energy dictate that the cell potential undergoes both Lorentz and β-gravitational transformations. The Lorentz transformation of cell potential results in slower rates of forward and backward redox reactions. Consequently, a moving observer will find the net current flow to be smaller. In a moving frame, the drop in potential requires a longer time to drive the same charge as in the rest frame of reference. This is fully consistent with relativistic time dilation, which predicts slower ticking of clocks for a moving observer. The decrease in cell potential for a moving observer is attributed to a reduction in the electromotive force operating between the ions and electrodes. This reduction in the magnitude of the electromotive force and cell potential is theoretically explained by the fusion of two spin-1 bosons (virtual photons) producing a spin-2 boson (graviton), which escapes into one of the possible $n$ extra dimensions of a total $4 + n$ dimensional spacetime. Similarly, the potential energy of nuclei under an applied magnetic field decreases for the same reason, namely enhanced graviton production in the moving frame. The gravitational transformation of cell potential also results in slower rates of forward and backward redox reactions. An observer in a stronger gravitational field will find the net current flow to be smaller in magnitude than an observer in a weaker gravitational field. This is because, in a stronger gravitational field, the drop in potential requires a longer time to drive the same charge compared to a weaker gravitational field. The decrease in cell potential in a stronger gravitational field is again attributed to the fusion of two spin-1 bosons (virtual photons) producing a spin-2 boson (graviton), which escapes into one of the possible $n$ extra dimensions of a total $4 + n$ dimensional spacetime. This leads to gravitational time dilation in electrode processes. Similarly, enhanced graviton production in a stronger gravitational field results in a decrease in the potential energy of nuclei under those conditions. The most significant result developed in this work is the Lorentz and gravitational transformation of electric potential, which establishes a connection between electromotive force and gravitational force through a thermodynamic state function. The present theory also offers a glimpse of physics beyond the event horizon of a black hole, a domain that remains largely unexplored. Since gravity is quantumphobic [10], and according to the β-gravitational transformation of energy spacing at the Schwarzschild radius, all quantization in a quantum molecular system is completely lost. Similarly, at the Schwarzschild radius, the β-gravitational transformation predicts that the electrical potential becomes zero due to the conversion of all virtual photons into gravitons, which subsequently escape into higher dimensions. When a molecule crosses the event horizon of a black hole, all its quantized states are destroyed because all virtual spin-1 particles combine to generate gravitons that escape into extra dimensions. The number of gravitons deposited on the event horizon as a molecule crosses it is uniquely determined by the molecule's ground-state electron density, analogous to the Hohenberg–Kohn theorem, which states that the ground-state electron density uniquely determines the external potential. At the event horizon, the electric potential effectively vanishes due to the β-gravitational transformation. As a consequence, particle–antiparticle pairs that spontaneously emerge from the vacuum, in accordance with the time–energy uncertainty principle, immediately lose their mutual electrostatic interaction upon creation. One particle is subsequently captured by the black hole, while its counterpart escapes to infinity, appearing as Hawking radiation [90]. This work seeks to expand the frontiers of black hole physics [91] by linking the passage of time to fluctuations of virtual particles in the fabric of spacetime. The present theory highlights the deep interconnections among quantum mechanics, relativity, and thermodynamics, while also suggesting that gravitons and extra dimensions are essential for explaining time dilation at the molecular scale. Thus, more physics [92] that is needed to study time dilation effects in molecular systems comes from M-Theory [23].

**Acknowledgements:** Special thanks to the late Prof. Sheng Hsien Lin for helpful discussions during the very early stages of this work in his research group, and for sponsoring a short scientific visit of Dr. M. W. Baig to Taiwan. Dr. M. W. Baig gratefully acknowledges Prof. Eli Pollak for his valuable and insightful explanation of the cross-over temperature.

**Funding information:** The authors state no funding involved.

**Author contribution:** The author has accepted responsibility for the entire content of this manuscript and approved its submission.

**Conflict of interest:** The authors state no conflict of interest.

**Data availability statement:** Data sharing is not applicable to this article as no datasets were generated or analysed during the current study.